# Synthesis and processing of lithium-loaded plastic scintillators on the kilogram scale

Michael J. Ford[a]*, Elisabeth Aigeldinger[a], Felicia Sutanto[a], Natalia P. Zaitseva[a], Viacheslav A. Li[a], M. Leslie Carman[a], Andrew Glenn[a], Cristian R. Catala[a], Steven A. Dazeley[a], Nathaniel Bowden[a]*

[a]*Lawrence Livermore National Laboratory*
*7000 East Avenue, Livermore, CA 94550*

**Abstract**

Plastic scintillators that can discriminate between gamma rays, fast neutrons, and thermal neutrons were synthesized and characterized while considering the balance between processing and performance at the kilogram scale. These trade-offs were necessitated by the inclusion of 0.1 wt. % lithium-6 to enable detection of thermal neutrons. The synthesis and processing of these plastic scintillators on the kilogram scale required consideration of many factors. First, a comonomer (methacrylic acid) was used to solubilize salts of lithium-6, which allow for a thermal-neutron capture reaction that produces scintillation light following energy transfer. Second, scintillation performance and processability were considered because the increasing content of the comonomer resulted in a sharp decrease in the light output. The use of small amounts of comonomer (≤3 wt. %) resulted in better performance but required high processing temperatures. At large scales, these high temperatures could initiate an exothermic polymerization that results in premature curing and/or defects. The deleterious effects of the comonomer may be mitigated by using m-terphenyl as a primary dye rather than 2,5-diphenyloxazole (PPO), which has been traditionally used in organic scintillators. Finally, the curing environment was controlled to avoid defects like cracking and discoloration while maintaining solubility of dopants during curing. For scintillators that were produced from kilogram-scale batches of precursors, the effective attenuation of scintillation light was characterized.

* This is to indicate the corresponding author.
  Email address: bowden20@llnl.gov; ford40@llnl.gov

## 1. Introduction

The rising cost of fossil fuels and concerns about greenhouse gas emissions motivate a revival of nuclear power generation.[1] However, the expansion of nuclear power throughout the world and the construction of novel reactor types may challenge the resources available for implementing conventional safeguards. Direct and nonintrusive measurements of reactor operation using offer one approach to address this concern.

These measurements can be facilitated by the development of novel detectors to enable near-field (ca. 10-100 m) monitoring of antineutrinos produced by a reactor.[2–10] Previous reports highlight the ability of organic scintillators to monitor this antineutrino flux. For example, the Precision Reactor Oscillation and Spectrum Experiment (PROSPECT)[11] recently demonstrated the measurement of the antineutrino spectrum from $^{235}$U at the High Flux Isotope Reactor at Oak Ridge National Laboratory. Like in many experiments that use organic scintillators to monitor antineutrino flux, PROSPECT utilized about 4 tons of a liquid scintillator loaded with a neutron capture agent ($^6$Li in this case), which can detect signatures of inverse beta decay. In this experiment, the scintillation light is associated with signatures of inverse beta decay using a measurement scheme called pulse-shape discrimination (PSD).[12] After monitoring a scintillation pulse over time, a prompt signal of scintillation light can be associated with the antineutrino energy, and the delayed signal of scintillation light can be associated with neutron capture by $^6$Li. Detectors like PROSPECT contain scintillator with this capability of PSD, making them useful for uniquely identifying capture reactions and rejecting fast-neutron background events.

For continued development of novel detectors, the phase of the detector material as it relates to the mobility of the detector may be considered. Liquid scintillators are relatively inexpensive, are easy to manufacture, and have good performance. However, liquid scintillators may require consideration of potential hazards (e.g., flammability), handling, and storage when used in mobile detectors. Conversely, plastic scintillators are less hazardous than liquid scintillators since they are in the solid state, and plastic scintillators are self-supporting, which is useful for mobility. However, plastic scintillators have not been widely available as materials capable of PSD until recently[13,14]. Additionally, the PSD performance of plastic scintillators worsens as the length of the scintillator increases due to light attenuation; this attenuation is detrimental to the performance of large-volume (ton-scale) detectors.[15] Thus, a mobile detector may require further consideration of these trade-offs and further development of detector materials like plastic scintillators.

One development that is important for plastic scintillators is the capability of these materials to discriminate between gamma rays, fast neutrons, and thermal neutrons (**Figure 1a**). This capability necessitates the inclusion of a neutron capture agent like $^{155}$Gd, $^{157}$Gd, $^{10}$B, or $^{6}$Li, and various reports describe attempts to incorporate neutron capture agents into plastic scintillators. In particular, the doping of scintillators with $^{6}$Li may be preferable since the $^{6}$Li(n,t)$\alpha$ capture reaction produces a localized, mono-energetic energy deposition that can be efficiently identified via PSD and energy selections.[16,17]

Carboxylate salts of $^{6}$Li have been previously used to incorporate $^{6}$Li into plastic scintillators while obtaining materials that are transparent. In one example, a $^{6}$Li salt of methacrylic acid (MAA) was copolymerized with styrene. The $^{6}$Li salt of MAA was not soluble in the plastic scintillator precursors at appreciable amounts; additional MAA was needed to dissolve the $^{6}$Li salt of MAA and increase the $^{6}$Li content. This necessity of additional MAA highlights how solubility of the polar $^{6}$Li compounds in the nonpolar matrix must be considered to produce plastic scintillators. Despite the need for additional MAA, these scintillators were promising for thermal-neutron detection. At a small scale ($\approx$ 1 cm), the plastic scintillator was responsive to an incident beam of thermal neutrons from a research reactor.[18] However, the optical attenuation properties were not assessed for this material but become important for large-volume detectors, where the longest side of a single scintillator may be on the order of 10-100 cm.

Other developments of these scintillators that contain $^6$Li focused on exploration of additional carboxylates.[19–22] One report described an investigation of 16 different $^6$Li salts. These $^6$Li salts were dissolved in a comonomer mixture of styrene and methacrylic acid (90:10 styrene:MAA) to determine the maximum solubility of $^6$Li at 60 °C. Additional $^6$Li could be dissolved as the MAA content increased, but the scintillation light output and figure of merit (FoM) for PSD decreased as MAA and $^6$Li content increased.[22] Thus, improvements in scintillation performance should consider the balance between processing and performance. This consideration is especially important in large plastic scintillators where thermal runaway[23] becomes a concern for processing, and attenuation becomes a concern for performance.

In this report, we describe the synthesis and processing of lithium-loaded plastic scintillators on the kilogram scale. We considered aspects related to the composition of the plastic scintillator like the primary dye, secondary dye, monomers, and lithium salts (**Figure 1b**) as well as aspects related to processing and curing like dissolution and temperature of cure (**Figure 1c**). We first synthesized various $^6$Li salts and characterized their solubility at different temperatures and with various concentrations of comonomers. Then, we considered trade-offs in processing and performance at the 10 g scale by evaluating the scintillation performance upon addition of comonomer and $^6$Li salt in plastic scintillators that contained 2,5-diphenyloxazole (PPO) as the primary dye. The light output of scintillation was reduced as the content of the comonomer MAA increased in these scintillators.

We then compared the performance of plastic scintillators that contain PPO vs. m-terphenyl (mTP) as the primary dye. The performance of plastic scintillators that contain mTP as the primary dye is less sensitive to the addition of MAA. When scaling our synthesis to 1 kg (**Figure 1d**), we targeted compositions that were available at scale and allowed for lower temperatures of processing to avoid thermally initiated polymerization and thermal runaway. All of these measures considered the potential for production of large-scale plastic scintillators, which will be useful for applications like mobile antineutrino detectors.

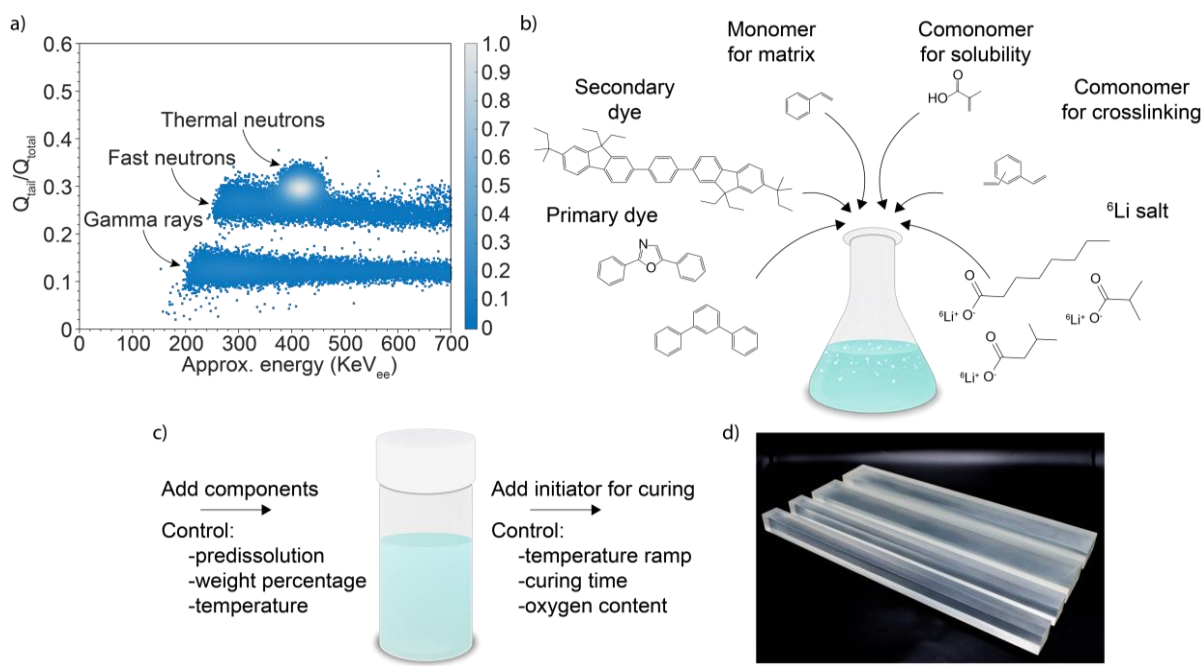

**Figure 1.** a) Plastic scintillators that are capable of PSD and contain lithium-6 distinguish between gamma rays, fast neutrons, and thermal neutrons, as shown in this PSD plot. The gradient scale bar represents a relative population of data points. b) The synthesis of plastic scintillators requires consideration of many components, as shown in this schematic. The chemical structures, starting from the bottom left and going clockwise, are those of m-terphenyl, 2,5-diphenyloxazole, Exalite 404, styrene, methacrylic acid, divinylbenzene, and three $^6$Li salts of carboxylic acids. c) The processing of the precursors requires control of dissolution and curing conditions. d) Control of synthesis and processing has enabled the production of large plastic scintillators, scaling to rectangular prisms like the ones in this photograph. The large scintillators are about 0.41 m in length (total mass of about 1.5 kg).

## 2. Materials and methods

Styrene (99%, Sigma), vinyl toluene (99%, Sigma), divinyl benzene (Sigma, technical grade), and methyl methacrylate (99%, VWR) were passed through an alumina column to remove inhibitor. Methacrylic acid (99%, Sigma) was dried over sodium chloride and distilled under vacuum to remove inhibitor. All monomers were sparged with nitrogen for > 30 min before being stored in a nitrogen-filled glovebox.

All monomers except methacrylic acid were stored in an inert atmosphere at −20 °C. Methacrylic acid was stored at room temperature. m-terphenyl (mTP, Smolecule) was purified by recrystallization from toluene. L-231 (Luperox, 1,1-di(t-butylperoxy)-3,3,5-trimethylcyclohexane) was used as a radical initiator after sparging for > 30 min with dry nitrogen and kept at −20 °C until needed. 2,5-diphenyloxazole (PPO, scintillation grade from Sigma), 1,4-bis(2-methylstyryl)benzene (bisMSB, Luxottica/Exciton), and 1,4-bis(9,9-diethyl-7-(*tert*-pentyl)-9*H*-fluoren-2-yl)benzene (E404, Luxottica/Exciton) were used as received without further purification.

$^6$Li carboxylate salts were synthesized by first suspending $^6$Li$_2$CO$_3$ (National Isotope Development Center) in a 1:1 mixture of water (deionized) and methanol (99.8%, Sigma). Excess carboxylic acid (1.02 equivalent excess) was mixed into a 1:1 mixture of water and methanol. The basic suspension was slowly added to the acid solution, and this mixture was heated to reflux for > 4 hours. The solution was filtered, and the $^6$Li salt was precipitated by adding excess volume of cold acetone. The $^6$Li salt was collected by vacuum filtration and washed with acetone, followed by drying under vacuum at 80 °C. This procedure was used for carboxylate salts of pentanoic acid, hexanoic acid, octanoic acid, 2-methylpropanoic acid, 2-methylbutanoic acid, 3-methylbutanoic acid, and 2-ethylhexanoic acid. All acids were purchased from Sigma or VWR and used as received.

Plastics were synthesized in a dry nitrogen environment. For initial evaluation, plastics were synthesized using 10 g of precursor materials. For the production of large plastic scintillators, the amount of precursors used was up to 2.7 kg. All materials that were not stored in a dry nitrogen environment were dried under vacuum. $^6$Li salts were easiest to process when they were first dissolved in a 1:1 mixture by weight of styrene:MAA at elevated temperatures (about 60-80 °C) before adding the remainder of the plastic composition. A typical synthesis would involve dissolving the primary dye (e.g., PPO) and the secondary dye (e.g., bis-MSB or E404) in styrene or vinyl toluene (VT). The monomer VT was used as the polymer matrix for plastics that used mTP as the primary dye. VT was used for these plastics as we observed less consistency in solubility for mTP-based plastics that used styrene. The performance of plastic scintillators that used styrene and VT were compared, and we observed no meaningful difference in performance for these plastics. This precursor solution was heated to about 60-80 °C and mixed with a solution of the $^6$Li salt in 1:1 mixture by weight of styrene:MAA. DVB (typically 5 wt. %) and initiator (0.08 wt. % for 10 g plastics; 0.01 wt. % for plastics batches greater than 400 g) were added. This mixture that contains the precursor solution mentioned earlier and the solution that contains $^6$Li was poured into a mould and sealed. The mould was placed in a nitrogen-filled oven and cured at elevated temperatures. In one experiment, the viscosity of the precursor solution was monitored using a rotary viscometer (Brookfield DV2T).

A typical curing profile would consist of heating for 7 days at 60 °C, followed by a temperature ramp to 75 °C over one day. The scintillators were cured in convection ovens (Cascade TEK) that were fitted with gas lines. Dry nitrogen flowed into solvent-resistance plastic bags that contained the mould inside the oven. The bags maintained a positive pressure of nitrogen. The plastic would stay at 75 °C for four days and then cool to room temperature over the course of one day. Following curing, the scintillators were removed from the moulds, then machined and polished. All photographs of samples were taken using a Nikon D750 and were globally edited in Adobe Lightroom for colour and exposure corrections.

For initial scintillator characterization, samples of mass equal to 10 g were measured. The outer edge and one face of the scintillators were wrapped and covered with Teflon tape. The exposed face was coupled with optical grease to a Hamamatsu R6231-100-SEL photomultiplier tube (PMT). Signals from the PMT were recorded at a sampling rate of 200 MS/s using a 14-bit CompuScope 14200 waveform digitizer. A relative quantification of light output (LO) was measured using ionizing radiation from $^{137}$Cs incident upon the plastic scintillator. The values of LO that we report in this manuscript are specific to our measurement system and thus should only be used for relative comparison. We normalized the value of LO to measurements of the commercial scintillator EJ-200. The location of 500 keVee was defined by the value of the pulse integral at 50% of the height of the $^{137}$Cs Compton edge. For many measurements, duplicate samples were synthesized, and averages are reported. For one condition, 9 samples were replicated in multiple batches and were measured. The standard deviation of these measurements was within 7% of the average value, which could be representative of the standard deviation related to contributions from measurement and synthesis. Where standard deviation is not reported, a conservative value of 10% of the value given could be assumed.

The measurement of effective attenuation length was performed with a longer scintillator bar (1″ x 1″ x 16″). This measurement employed a setup that was identical to the setup used to characterize scintillator bars in an antineutrino detector called SANDD (Segmented AntiNeutrino Directional Detector).[24] This bar was wrapped with polytetrafluoroethylene tape (POLY-TEMP PN-16050), and a pair of 1" Hamamatsu R1924A-100 PMTs were mounted at either end of the scintillator bar using EJ-550 silicone optical grease. The PMT operating voltage was set at -1100 V, and signals were digitized using a Struck SIS3316 digitizer module (250 MS/s, 14 bit, 5 V dynamic range). The energy threshold was set at approximately 0.1 MeV$_{ee}$, and 1600 ns-long waveforms were sent to disk and stored in ROOT data format. The charge response difference between the two PMTs due to gain and optical coupling variation was corrected using a collimated $^{137}$Cs gamma-ray source directed at the center of the plastic bar. Here, lead bricks were used for shaping the gamma-ray source into a fan beam of about 0.5 cm width.

To obtain the effective attenuation length of the scintillator bar, we performed a series of collimated $^{22}$Na measurements at regular intervals along the length of the bar. In each measurement, we identified the location of the Compton continuum maximum of the 1.275 MeV gamma-ray by fitting the energy response with a Gaussian profile while varying the range of the fit to find the minimum $\chi^2$ (best fit). The Compton continuum maximum position was identified as the mean of the Gaussian profile that yielded the minimum $\chi^2$. The associated uncertainty was estimated by varying the range of the fit until the $\chi^2$ value exceeded the 68% confidence level (CL) of the minimum $\chi^2$; the uncertainty was the corresponding range of the mean of the Gaussian profile.

To measure PSD in smaller plastics, plastic scintillators were exposed to a $^{252}$Cf source. The source was shielded behind 5.1 cm of lead to reduce the gamma-ray flux. To obtain a flux of thermal neutrons, high density polyethylene was also used as a moderator for $^{252}$Cf. The measurements of scintillation from plastic scintillators exposed to $^{252}$Cf were integrated over time to determine the total charge ($Q_{total}$). The charge of the delayed component of the signal ($Q_{tail}$) was determined from a delayed fraction of the scintillation pulse. Scintillation pulses due to interactions of the scintillator with neutrons have a larger fraction of $Q_{tail}$ relative to $Q_{total}$; therefore, a comparison of $Q_{tail}$ relative to $Q_{total}$ can be used to distinguish between scintillation due to neutrons vs. gamma rays. The PSD was quantified using a figure of merit (FoM) that is determined from histograms of the ratio of the charge of the delayed component relative to the total charge, as described in previous reports[15,25]. Briefly, the FoM is:

$$FoM = \frac{\langle n,t \rangle - \langle \gamma \rangle}{FWHM_{n,t} + FWHM_\gamma} \qquad (2)$$

In this equation, $\langle n,t \rangle - \langle \gamma \rangle$ represents the difference between the average value of the neutron and gamma-ray signals, and $FWHM_{n,t} + FWHM_\gamma$ represents the sum of the full-width at half of the maximum value of the distributions of the thermal-neutron and gamma-ray signals at the electron-equivalent energy of the thermal-neutron spot. For plastics that don't contain $^6$Li, the same equation was used for FoM, but the position and FWHM of the neutron peak is used at an electron-equivalent energy near the $^{137}$Cs Compton edge.

For PSD of the larger scintillator bar (1″ × 1″ × 16″), an identical setup as the measurement of effective attenuation length was used. The bar was irradiated with an uncollimated $^{252}$Cf source, and lead bricks with a total thickness of 6″ were placed between the detector and the $^{252}$Cf source to reduce the gamma-ray flux. The charge integration limits were optimized, and the best parameters were found to be [$t_L$-20 ns ≤ $Q_{total}$ ≤ $t_L$+1300 ns] and [$t_L$+24 ns ≤ $Q_{tail}$ ≤ $t_L$+1300 ns], where $t_L$ is the leading edge of the waveform. Assuming light transport behaves exponentially along the length of the scintillator bar, we can eliminate the dependence of energy on event position by reconstructing the energy as $E = \sqrt{E_A E_B}$, where $E_A$ and $E_B$ are the charges collected by the two PMTs.

## 3. Results and discussion

*3a. Selection of lithium-6 salt based on solubility at moderate temperatures*

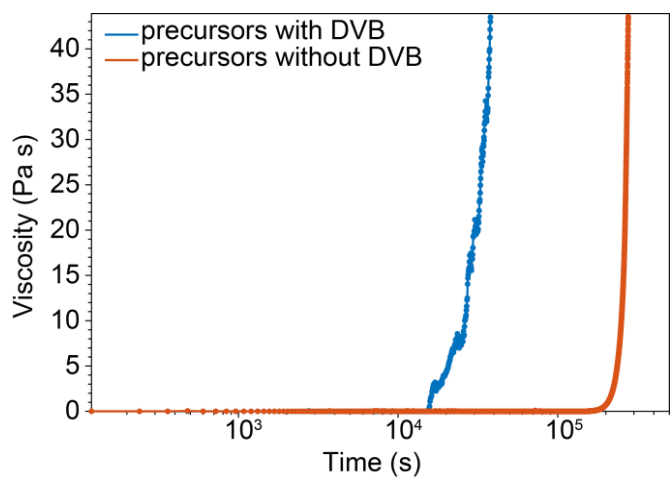

**Figure 2**. The viscosity of liquid precursors containing 30 wt. % PPO and 0.2 wt. % bis-MSB can increase over time as polymerization occurs, and this increase in viscosity prevents processing of the liquid into a mould. In this case, precursors with 8 wt. % DVB and without DVB are compared while held at a temperature of 50 °C.

To make large-scale production of plastic scintillators easier, our selection of materials focused on those materials that had simple processing requirements (e.g., temperatures below 80 °C). This requirement is necessary since high processing temperatures for large plastic scintillators could thermally initiate the polymerization. After dissolution of all dopants, we monitored the viscosity over time of liquid precursors with and without the crosslinker divinylbenzene (DVB) at a temperature of 50 °C (**Figure 2**). The liquid precursors did not contain a radical initiator that initiates polymerization; thus, any increase in viscosity is due to thermally initiated polymerization. For precursor liquids that contained DVB, the viscosity began to increase to measurable values above 0.1 Pa s after about 15600 s (4.3 hours) at 50 °C. The viscosity further increased, reaching values greater than 40 Pa s after about 37400 s (10.4 hours) at 50 °C. While these timescales may be appropriate to process plastic scintillators with industrial equipment, we observed that complete dissolution of all components often required > 12 hours of stirring at elevated temperatures.

The premature onset of polymerization caused this increase in viscosity, which could prevent trapped air bubbles from escaping or even prohibit transfer of the liquid precursor to a mould. For precursor liquids that did not contain DVB, the viscosity began to increase to measurable values above 0.1 Pa s after about 175000 s (48.6 hours) at 50 °C. As before, the viscosity further increased and reached values greater than 40 Pa s after 275000 s (76.4 hours) 50 °C. Without DVB, the working time of the liquid precursors can be increased. Still, the limited working time of these materials highlights the need for simple processing requirements like low temperatures.

**Table 1.** Summary of solubility tests. Note that some salts that formed a gel phase were initially soluble. For solubility tests at 23 °C, the solubility was observed after 20 hours of mixing. For solubility tests at 65 °C, the solubility was observed after 30 minutes at 65 °C following 2 hours at 50 °C.

| Acid used for $^6$Li salt | Solubility, 23 °C, 85:15 styrene:MAA | Solubility, 65 °C, 85:15 styrene:MAA |
|---|---|---|
| Pentanoic acid | Insoluble | Soluble |
| Hexanoic acid | Insoluble | Soluble |
| Octanoic acid | Insoluble | Soluble |
| 2-methylpropanoic acid | Formed gel | Soluble |
| 2-methylbutanoic acid | Formed gel | Soluble |
| 3-methylbutanoic acid | Soluble | Soluble |
| 2-ethylhexanoic acid | Formed gel | Soluble |

Another requirement for simple processing relates to the relative solubility of the $^6$Li salts. Thus, we evaluated the solubility of various carboxylate salts of $^6$Li while considering the need for simple processing requirements. We synthesized $^6$Li salts of pentanoic acid, hexanoic acid, octanoic acid, 2-methylpropanoic acid, 2-methylbutanoic acid, 3-methylbutanoic acid, and 2-ethylhexanoic acid (**Figure 3a**). The synthesis and solubility of these salts have been described previously[22], but our focus is on solubility for synthesis of large plastics, which requires further considerations related to processing that have not been reported. We added the $^6$Li salts to liquid precursors that contained all monomers and dopants. The composition of the liquid precursor was as follows: 30 wt. % 2,5-diphenyloxazole (PPO); 0.2 wt. % 1,4-bis(2-methylstyryl)benzene (bisMSB); 5 wt. % DVB; an equivalent amount of $^6$Li salt to obtain 0.1 wt. % $^6$Li; and the remainder was a mixture of 85 wt. % styrene and 15 wt. % methacrylic acid (MAA). The MAA is necessary to dissolve the $^6$Li salt; other monomers like methyl methacrylate and methyl acrylate do not dissolve the $^6$Li salt.

All precursors were mixed into a single vial and allowed to equilibrate for 20 hours at room temperature (23 $^o$C). Lithium-6 2-methylbutanoate was readily soluble in the mixture initially. However, an opaque gel formed within 2 hours and persisted after 20 hours (**Figure 3b**). Lithium-6 2-ethylhexanoate also formed a gel after initial dissolution. For lithium-6 2-methylpropanoate, an opaque gel was observed after 20 hours, but this mixture never fully dissolved, suggesting low solubility of this $^6$Li salt. Similarly, $^6$Li salts of the linear alkyl carboxylic acids (pentanoic acid, hexanoic acid, and octanoic acid) never fully dissolved at 23 $^o$C. For lithium-6 3-methylbutanoate, the liquid precursor remained clear after 20 hours at 23 $^o$C.

The opaque gels that we observed could be destabilized after heating to elevated temperatures. All vials were heated to 50 $^o$C for 2 hours, which improved dissolution of all components that were insoluble at room temperature. For example, the liquids that contained $^6$Li salts of 2-methylbutanoic acid and 2-ethylhexanoic acid became transparent. Further heating to 65 $^o$C for 30 minutes improved dissolution; all liquid precursors with different $^6$Li salts were transparent after this heating step except for the precursor that contained lithium-6 2-methylpropanoate (**Figure 3c**).

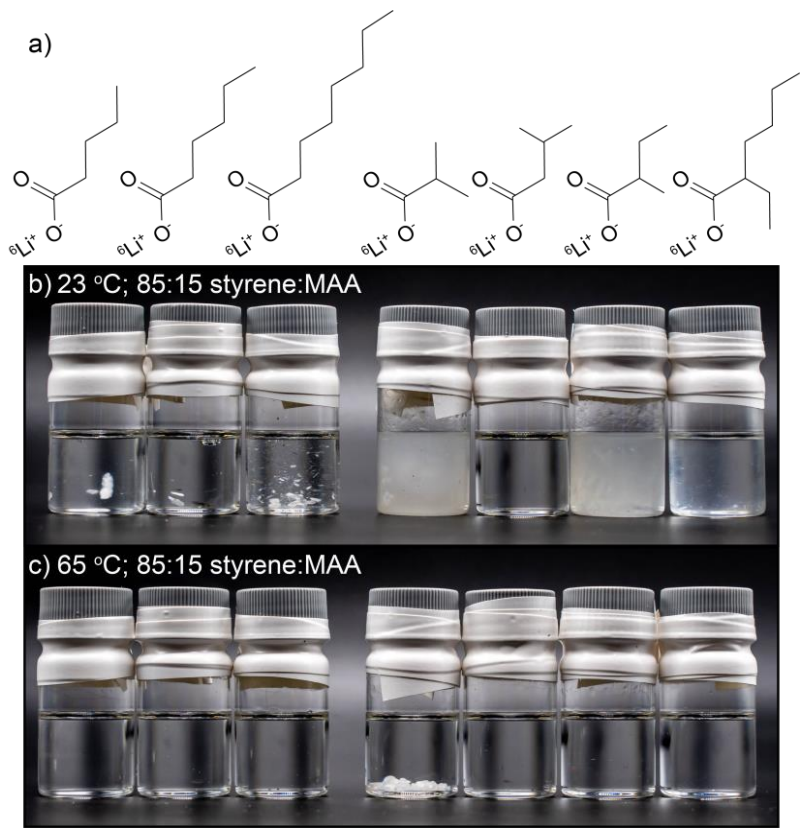

**Figure 3**. a) Chemical structures of $^6$Li salts that were studied; from left to right the structures correspond to $^6$Li salts of pentanoic acid, hexanoic acid, octanoic acid, 2-methylpropanoic acid, 2-methylbutanoic acid, 3-methylbutanoic acid, and 2-ethylhexanoic acid. b) Photographs of liquid precursors in vials following equilibration for 20 hours at 23 °C. c) Photographs of liquid precursors in vials following equilibration at an additional 2 hours at 50 °C plus 30 minutes at 65 °C. All liquid precursors contain all components of a plastic scintillator except the radical initiator. The precursors differ in the $^6$Li salt that was added; from left to right, the vials contain $^6$Li salts of pentanoic acid, hexanoic acid, octanoic acid, 2-methylpropanoic acid, 3-methylbutanoic acid, 2-methylbutanoic acid, and 2-ethylhexanoic acid. The vials are 28 mm in diameter.

Based on this analysis, we selected the $^6$Li salt of 3-methylbutanoic acid for the synthesis of large plastics; however, other $^6$Li salts like lithium-6 pentanoate and lithium-6 2-methylbutanoate may also be suitable given that they form transparent precursors after dissolution at 65 °C. Furthermore, we sometimes observed the formation of a gel phase for precursors that contained lithium-6 3-methylbutanoate when using less MAA, which highlights how plastic scintillators that contain this $^6$Li salt are not immune to this processing challenge.

To avoid the formation of this gel phase, the $^6$Li salt could be dissolved separately from the rest of the dopants. A 1:1 mixture of styrene and MAA was sufficient to avoid thermally initiated homopolymerization of MAA during dissolution; homopolymerization of MAA results in an opaque material. The remainder of the styrene needed to form the final plastic was used to dissolve the primary and secondary dye in a separate container. Then, the two separate mixtures could be heated to 60-80 °C and mixed at elevated temperatures before adding DVB and the radical initiator and casting in a mould.

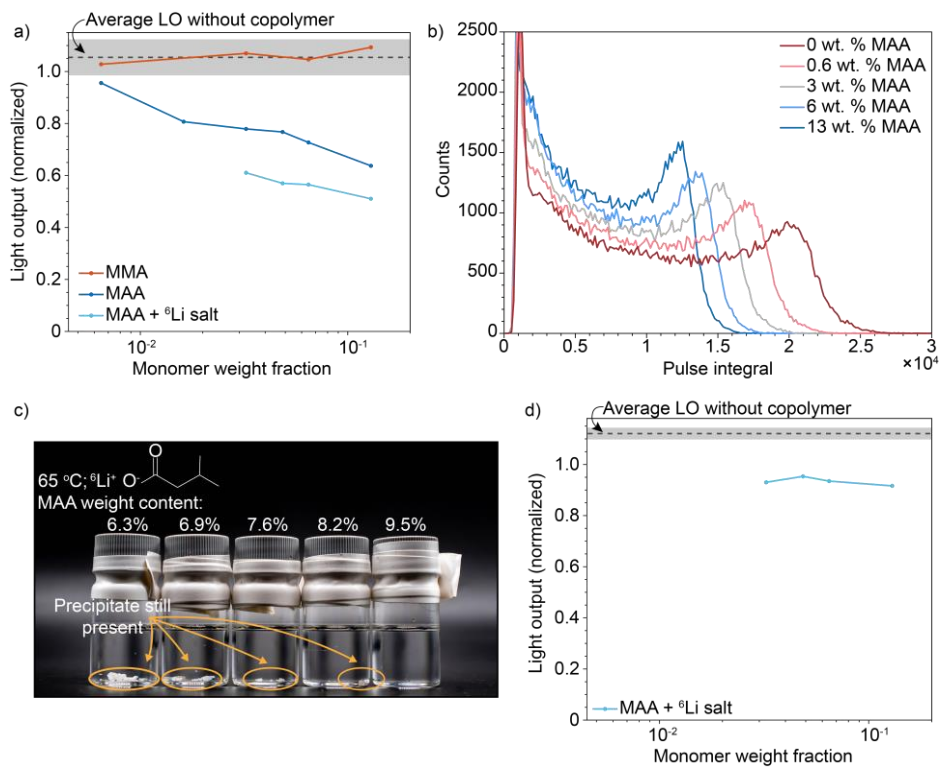

**Figure 4**. a) A detrimental effect of MAA and $^6$Li salts reduce light output (LO) of scintillators that contain PPO as a primary dye. $^6$Li salt may also influence light output. b) The LO decreases as the content of MAA increases, which can be observed in the histograms that show the $^{137}$Cs Compton edge. c) The photograph of these vials highlights the solubility threshold of lithium-6 3-methylbutanoate at 65 °C. The vials are 28 mm in diameter. d) Less substantial effect on LO by MAA and $^6$Li salt for scintillators that contain mTP instead of PPO. The $^6$Li salt used for all samples referenced in this figure was lithium-6 3-methylbutanoate. For a) and b), the average (dashed line) and standard deviation (grey shaded region) of scintillators that do not contain any comonomer are shown for reference.

### 3b. Effect of comonomer on scintillation performance

Importantly, the effect of the comonomer that solubilizes the $^6$Li salts on scintillation performance should be evaluated. The addition of non-aromatic comonomers like methyl methacrylate (MMA) or methacrylic acid (MAA) can reduce scintillation performance.[22,26] When 26 wt. % of MMA was used in a plastic scintillator, the light yield reduced by 5% when compared to a plastic scintillator that contained only polystyrene as the matrix. When 58 wt. % of MMA was used in a plastic scintillator, the light yield reduced by 18%.[26] The total amount of the comonomer

MAA added for dissolution of $^6$Li salts is typically less than 20 wt. % of the total material, so we instead focused on comonomer addition at these lower concentrations.

**Table 2.** Summary of effect of composition on light output.

| Primary dye | Co-monomer | Co-monomer content | Lithium salt content | Light output |
|---|---|---|---|---|
| PPO | N/A | 0 | 0 | 1.05 |
| mTP | N/A | 0 | 0 | 1.12 |
|  |  |  |  |  |
| PPO | MMA | 0.6 | 0 | 1.03 |
| PPO | MMA | 3 | 0 | 1.07 |
| PPO | MMA | 6 | 0 | 1.05 |
| PPO | MMA | 13 | 0 | 1.09 |
|  |  |  |  |  |
| PPO | MAA | 0.6 | 0 | 0.96 |
| PPO | MAA | 2 | 0 | 0.81 |
| PPO | MAA | 3 | 0 | 0.78 |
| PPO | MAA | 5 | 0 | 0.77 |
| PPO | MAA | 6 | 0 | 0.73 |
| PPO | MAA | 13 | 0 | 0.64 |
|  |  |  |  |  |
| PPO | MAA | 3 | 1.7 | 0.61 |
| PPO | MAA | 5 | 1.7 | 0.57 |
| PPO | MAA | 6 | 1.7 | 0.56 |
| PPO | MAA | 13 | 1.7 | 0.51 |
|  |  |  |  |  |
| mTP | MAA | 3 | 1.7 | 0.93 |
| mTP | MAA | 5 | 1.7 | 0.95 |
| mTP | MAA | 6 | 1.7 | 0.93 |
| mTP | MAA | 13 | 1.7 | 0.92 |

Even though MMA does not solubilize $^6$Li salts that we studied, we used this comonomer as a non-aromatic additive to compare its effect on performance to scintillators that contain the solubilizing comonomer, MAA. We also compared the performance of plastic scintillators that did not contain any comonomer. The average light output (LO) of three separate samples that did not contain any comonomer was 1.05 with a standard deviation of 0.07 (**Figure 4a, Table 2**). With 0.6 wt. % MMA added, the LO was 1.03. As the amount of MMA increased, there was no clear trend in the LO. At 13 wt. % MMA, the LO was 1.09, and all measured values of LO were within a standard deviation of the average value of plastic scintillators that did not contain a comonomer. Thus, at these concentrations of MMA comonomer, the energy transfer and light emission do not appear to be affected.

The same trend did not persist when using MAA. At 0.6 wt. % MAA, the LO was 0.96, which corresponds to a 9% reduction in LO when compared to plastic scintillators without this comonomer. At 1.6% MAA, the LO was further reduced to 0.81, which is a 23% reduction. The LO continues to decrease as MAA content increases (**Figure 4b, Table 2**), but the magnitude of reduction appears to taper as the MAA content exceeds 3 wt. %. At 13 wt. % MAA, the LO was reduced to 0.64, which corresponds to a 40% reduction. The discrepancy between the effects of MMA and MAA on LO indicates that the decrease in LO upon addition of MAA does not result from simple dilution by a non-aromatic material. Rather, this decrease in LO suggests that MAA may be detrimental to processes that affect scintillation like energy transfer or emission. Although the exact mechanism is not fully elucidated, it is possible that the heteroatoms on PPO (N, O) may interact with the polar acid functional group on MAA.

Notably, the LO further decreases upon addition of $^6$Li salts to plastics that contain PPO as the primary dye along with MAA as a comonomer when compared to the addition of MAA alone. At 13 wt. % MAA, the LO decreased from 0.64 to 0.52 when adding the $^6$Li salt. The reduction in LO with increasing MAA content and addition of $^6$Li salt poses a challenge related to processing: higher content of MAA allows for processing of the plastic scintillator at lower temperatures but reduces LO. Plastic scintillators with lower content of MAA are more difficult to process due to poor solubility of the $^6$Li salt. For example, lithium-6 3-methylbutanoate is not fully soluble at 65 °C when the MAA content is equal to or less than 8 wt. % (**Figure 4c**). The cause of the reduction in performance upon addition of $^6$Li salt may be similar to the reduction in performance upon addition of MAA; PPO may have unfavourable interactions and/or reactivity with these polar molecules.

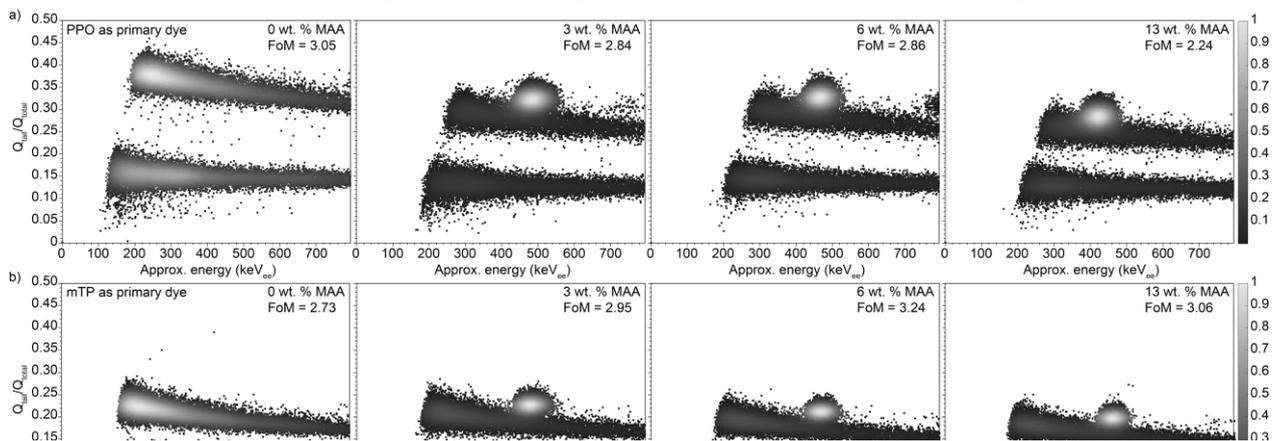

**Figure 5.** PSD distributions used to calculate FoM for comparison of plastics that contain PPO (a) and mTP (b). Note that the FoMs for $^6$Li-plastics are given for discrimination between thermal neutrons and gamma rays whereas FoMs for plastics without $^6$Li correspond to discrimination between fast neutrons and gamma rays.

To test this idea, we compared the scintillation performance when using 30 wt. % m-terphenyl (mTP) as a primary dye instead of PPO. The chemical structure of mTP (**Figure 1b**) does not contain any heteroatoms (i.e., only contains C, H). Plastics that used mTP as the primary dye but did not contain MAA or a $^6$Li salt had an average LO of 1.12 (**Figure 4d**, **Table 2**). Plastics that contained MAA and $^6$Li had a slight reduction in LO with values of 0.93, 0.95, 0.93, and 0.92 at 3 wt. %, 5 wt. %, 6 wt. %, and 13 wt. % MAA. At 13 wt. % MAA, the LO of a plastic that contains $^6$Li was reduced by 18% when compared to a plastic that contained no MAA or $^6$Li. This value of LO is nearly double the LO of an equivalent plastic that contained PPO instead of mTP as the primary dye.

We also compared the FoM for PSD. For plastics that contain PPO but no $^6$Li, the FoM is 3.05, which provides a baseline for comparison after addition of MAA and $^6$Li. Note that this FoM compares discrimination between gamma rays and fast neutrons whereas the FoM for plastics that contain $^6$Li compares discrimination between gamma rays and thermal neutrons. The energy range used to determine FoM was 450-550 keV$_{ee}$, but that range was adjusted to capture the thermal neutron peak. Upon addition of 13 wt. % MAA and $^6$Li, the FoM decreases to 2.24 and decreases for all plastics that were measured that contained MAA (**Figure 5a**).

These PSD results can also be compared to plastics that contain mTP instead of PPO (**Figure 5b**). For plastics that contain mTP but no $^6$Li, the FoM is 2.73. Upon addition of MAA and $^6$Li, the FoM is between 2.95 and 3.24. This increase is mostly attributed to the high value of $Q_{tail}/Q_{total}$ for the thermal-neutron capture spot along with the observation that plastics that contain mTP are less sensitive to the addition of the polar compounds that enable thermal-neutron capture. The FoM for plastics that contained PPO decreased when MAA and $^6$Li were added whereas the FoM for plastics that contained mTP increased when MAA and $^6$Li were added.

### 3c. Production of large plastic scintillators

As shown above, mTP may be promising for high performance scintillators that contain $^6$Li; however, the inconsistencies with solubility and the current availability at large scales in sufficient purity from commercial suppliers make mTP less suitable for the production of large plastic scintillators. For these large plastic scintillators, we selected PPO despite its lower LO and lower FoM at smaller scales.

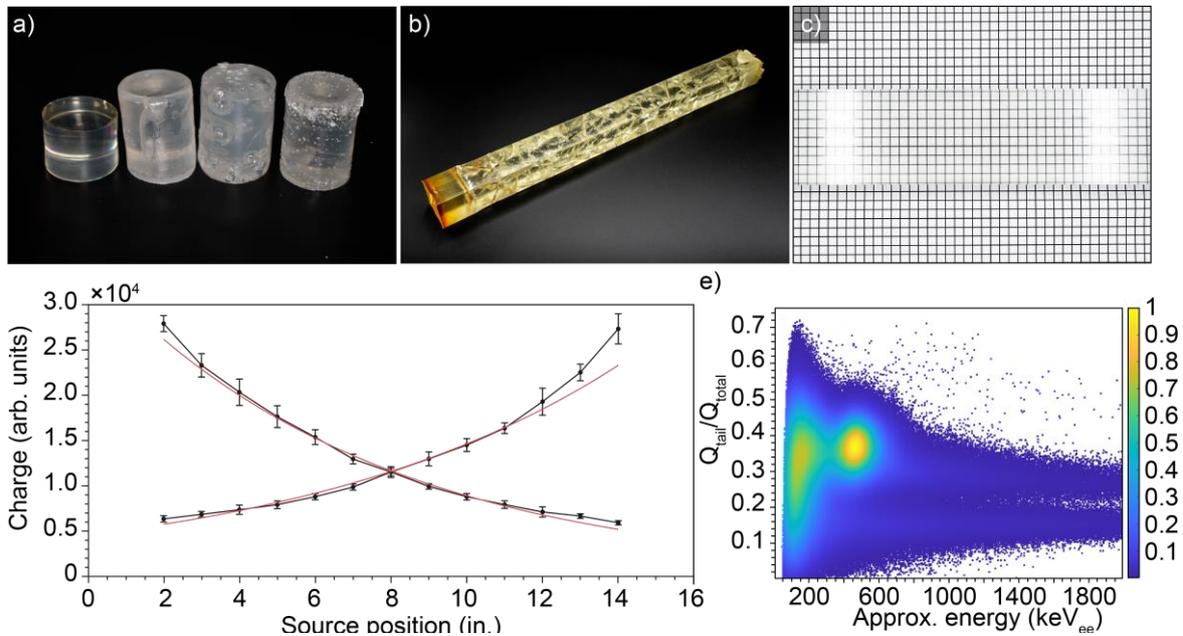

**Figure 6.** a,b) Conditions that are not optimized for the production of large plastics produce defects like cracks and bubbles (a) and discoloration (b). The left-most scintillator in (a) does not contain defects and serves as a reference. These scintillators in (a) have a diameter of about 5.5 cm. The plastic scintillator in (b) is 40 cm in length. c) When the curing conditions and environment are controlled, plastics scintillators can be produced in large scale, as shown in this photograph of a scintillator that is 5 cm in width atop a sheet of paper. d) The effective attenuation length was measured by placing a collimated gamma-ray source at set distances away from two PMTs and measuring the PMT response (black). The data were fitted with an exponential profile to estimate the effective attenuation length (19-21cm). e) Large plastic scintillators are capable of PSD, as shown in this distribution that demonstrates an ability to separate signals from thermal neutrons and gamma rays.

Similarly, other secondary dyes like Exalite 404 (E404) may have the best performance in the plastic scintillators that we evaluated[27], but the cost of E404 might be prohibitive when compared to a secondary dye like 1,4-bis(2-methylstyryl)benzene (bisMSB).

When processing the plastics, all precursors except DVB and the radical initiator are slowly heated to temperatures between 60 and 80 °C until full dissolution. The plastics are cured in glass or aluminium moulds. To control the rate of polymerization, the radical initiator is added at concentrations of 0.01 wt. %, and plastics are cured at an initial temperature of 60 °C. Various times of curing were used, and a typical recipe would involve curing for 7 days at 60 °C followed by an additional 4 days of curing at 75 °C. Excessive radical concentration and/or heating during processing and curing could lead to defects like cracks and bubbles (**Figure 6a**). Precipitation of precursors that have lower solubility may occur if temperatures are too low. When curing the plastics, oxygen is displaced by a steady flow of nitrogen; without nitrogen flow, discoloration can occur (**Figure 6b**). These precautionary measures allow us to produce plastic scintillators without defects and with minimal discoloration (**Figure 6c**). The effective attenuation of a plastic scintillator that was 16" long was measured by placing a collimated gamma-ray source near the scintillator and measuring the total light detected by PMTs that are mounted on each end of the plastic scintillator (**Figure 6d**). The effective attenuation was determined to be about 19-21 cm, which is comparable to the value obtained in previous experiments.[24] This plastic also had PSD capability; a thermal-neutron spot is clearly separated from the gamma signal (**Figure 6e**). After we optimized our process for synthesis of these large plastics, we outsourced production to Eljen Technologies who is currently producing $^6$Li-loaded prototypes with dimensions exceeding 0.5 m; full characterization of scintillator performance at these large scales will be the subject of a future publication.

## 4. Conclusion

By careful control of composition and processing, plastic scintillators that can discriminate between gamma rays, fast neutrons, and thermal neutrons can be produced at a scale of 1 kg or greater. The solubility of dopants that enable scintillation functionality and solubilizing additives like methacrylic acid (MAA) that may negatively affect performance were considered. Synthesis and processing procedures were developed for large plastic scintillators containing 0.1 wt. % $^6$Li and high concentration (30 wt. %) of PPO used as a primary dye. These scintillators were capable of PSD. In these studies, various $^6$Li salts of aliphatic carboxylic acids were evaluated, and many were found to be suitable for the production of large plastic scintillators with the addition of MAA. The amount of MAA that was added to solubilize $^6$Li salts affected the scintillation performance but also determined the temperature that plastic scintillators could be produced at. An alternative way to avoid the deleterious effects of MAA was discovered; use of m-terphenyl instead of PPO improved plastic scintillators. However, m-terphenyl may have limitations like availability at large volumes.

With these considerations in mind, methods for the preparation of plastic scintillators loaded with $^6$Li were established and demonstrated. Large-volume pieces that could be used for large detectors were produced.[24,28] Such detectors will be important for future safeguards related to nuclear power production and for unravelling unknown aspects of particle physics.

## Acknowledgements

This work was performed under the auspices of the U.S. Department of Energy by Lawrence Livermore National Laboratory under Contract DE-AC52-07NA27344 and was supported by the LLNL-LDRD Program under Project No. 20-SI-003, release number LLNL-JRNL-839909. We would like to thank Jacob Kim for careful reading and discussion of this manuscript.

The authors declare no competing interests.